\documentclass[12pt]{article}
\begin{document}
\newcommand{\nc}[2]{\newcommand{#1}{#2}}
\newcommand{\ncx}[3]{\newcommand{#1}[#2]{#3}}
\ncx{\dd}{2}{\frac{\partial #1}{\partial #2}}
\nc{\cD}{{\cal D}}
\nc{\cJ}{{\cal J}}
\nc{\cQ}{{\cal Q}}
\nc{\ag}{\alpha}
\nc{\Gam}{\Gamma}
\nc{\del}{\delta}
\nc{\eps}{\epsilon}
\nc{\Th}{\Theta}
\nc{\lb}{\lambda}
\nc{\ps}{\psi}
\nc{\og}{\omega}
\nc{\lh}{\left(}
\nc{\rh}{\right)}
\nc{\ld}{\left.}
\nc{\rd}{\right.}

\def\R{{\bf R}}
\def\d{{\rm d}}
\def\p#1{{\partial}_{#1}}

\ncx{\abs}{1}{\left| #1 \right|}

\title{f-symbols in Robertson-Walker space-times}

\author{Florian Catalin Popa \thanks{E-mail:~~~catalin@venus.nipne.ro} and
Ovidiu Tintareanu-Mircea \thanks{E-mail:~~~ovidiu@venus.nipne.ro}\\
{\small \it Institute of Space Sciences}\\
{\small \it Atomistilor 409, RO 077125, P.O.Box M.G.-23, Magurele, Bucharest, Romania}}

\date{\today}

\maketitle

\begin{abstract}
In a Robertson-Walker space-time a spinning particle model is investigated
and we show that in a stationary case, there exists a class of new structures
called {\em f-symbols} which can generate reducible Killing tensors and supersymmetry algebras.

~

Keywords: {\it f-symbols, Killing-Yano tensors, hidden symmetries}

~

Pacs: 04.20.-q, 12.60.Jv, 04.65.+e

\end{abstract}

\section{Introduction}\label{sec-intro}
The constants of motion of a scalar particle in a curved space are
determined by the symmetries of the manifold, and are expressible in
terms of the Killing vectors and tensors. A similar result hold for
a spinning particle, with the modification that the constants of motion
related to a Killing vector contain spin dependent parts and, there are
Grassmann odd constants of motion, which do not have a counterpart
in the scalar model. An illustration of the existence of extra conserved
quantities is provided by Kerr-Newmann \cite{Carter} and Taub-NUT geometry. For
the geodesic motion in the Taub-NUT space,the conserved vector analogous
to the Runge-Lenz vector of the Kepler type problem is quadratic in $4$-velocities,
its components are Stackel-Killing tensors and they can be expressed as symmetrized
products of Killing-Yano tensors \cite{Ya, GiRu1, GiRu2, VaVi1, vH}.
The configuration space of spinning particles (spinning space) is
an extension of an ordinary Riemannian manifold, parameterized by
local coordinates $\left\lbrace x^\mu \right\rbrace$, to a graded
manifold parameterized by local coordinates $\left\lbrace
x^\mu,\psi^\mu\right\rbrace$, with the first set of variables
being Grassmann-even (commuting) and the second set Grassmann-odd
(anti-commuting). In the spinning case the generalized Killing equations are more
involved and new procedures have been conceived.

The aim of this paper is to show that the existence of the Killing-Yano tensors and
its properties in a stationary case of Robertson-Walker space-times,
may be understood in a systematic way as a particularly interesting example
of a more general structures called $f$-symbols \cite{Gibbons}.

The plan of this paper is as follows.
In Sec. \ref{sec-symm} we give a short review of the formalism for
pseudo-classical spinning point particles in an arbitrary background space-time,
using anticommuting Grassmann variables to describe the spin degrees
of freedom.
In Sec. \ref{sec-new-symm} we present the properties of $f$-symbols
and the role played in generating new supersymmetries.
In Sec. \ref{sec-f-symb} we apply the results previously presented
to show that in a stationary case of a Robertson-Walker space-time
we obtain explicit solutions for $f$-symbols equations and construct corresponding
Killing tensors.
Conclusions are presented in Sec. \ref{sec-concl}.

\section{Symmetries of Spinning Particle Model}\label{sec-symm}

The symmetries of spinning particle model can
be divided into two classes \cite{Ri, RivH}: {\it generic} ones, which exists
for any spinning particle model and {\it non-generic} ones, which
depend on the specific background space considered. To the first
class belong proper-time translations and supersymmetry, generated
by the hamiltonian and supercharge: $ Q_0=\Pi_\mu \psi^\mu. $ To
obtain all symmetries, including the non-generic ones, one has to
find all functions ${\cal J}(x,\Pi,\ps)$ which commute with the
Hamiltonian in the sense of Poisson-Dirac brackets: $ \left\{ H,
\cJ \right\}\, =\, 0.$
Expanding ${\cal J}$ in a power series in the covariant momentum

\begin{equation}
{\cal J}\, =\, \sum_{n=0}^{\infty}\, \frac{1}{n!}\,
J^{(n)\, \mu_{1}...\mu_{n}}(x,\ps)\, \Pi_{\mu_{1}}\, ... \Pi_{\mu_{n}},
\label{3.5}
\end{equation}
then the components of ${\cal J}$ satisfy
\begin{equation}
D_{\left( \mu_{n+1} \right.}\, J^{(n)}_{\left. \mu_{1}\dots\mu_{n}
\right)}\,
   +\, \og_{\lh \mu_{n+1} \rd \:\:b}^{\:\:\:\:a}\, \ps^{b}\,
   \dd{J^{(n)}_{\ld \mu_{1}\dots\mu_{n} \rh}}{\ps^{a}}
   =\, R_{\nu \left(\mu_{n+1}\right. }\, J^{(n+1)\:\:\:\:\nu}_{\left.
          \mu_{1}... \mu_{n} \right) },
\label{3.6}
\end{equation}
where the parentheses denote full symmetrization over
the indices enclosed, $\og_{\mu\:\:b}^{\:\:a}$ the spin connection and $R_{\mu\nu}$
given by
\begin{equation}
R_{\mu\nu}  =  { \frac{i}{2}\, \ps^{a} \ps^{b} R_{ab \mu\nu}. }\\
\end{equation}
Eqs.(\ref{3.6}) are the generalizations of
the Killing equations to spinning space first obtained in \cite{Ri}.
Writing for ${\cal J}$ the series expansion
\begin{equation}
{\cal J}(x,\Pi,\ps)\, =\, \sum_{m,n = 0}^{\infty}\,
\frac{i^{\left[ \frac{m}{2}
   \right]}}{m!n!}\, \ps^{a_{1}}...\ps^{a_{m}}\, f^{(m,n)\,
\mu_{1}...\mu_{n}}_{
   \: a_{1}...a_{m} }(x)\, \Pi_{\mu_{1}}...\Pi_{\mu_{n}},
\label{3.11.1}
\end{equation}
where $f^{(n,m)}$ is completely symmetric in the $n$
upper indices $\{\mu_{k}\}$ and completely anti-symmetric in the
$m$ lower indices $\{a_{i}\}$ we obtains the component equation
\begin{equation}
n\, f^{(m+1,n-1)\, \lh \mu_{1}...\mu_{n-1} \right.}_{\: a_{0}
a_{1}...a_{m}}\,
  e^{\left. \mu_{n} \right)\, a_{0} }\, =\, m\, D_{\left[ a_{1} \right.}\,
  f_{\left. \: a_{2}...a_{m} \right]}^{(m-1,n)\, \mu_{1}...\mu_{n}} ,
\label{3.11.2}
\end{equation}
where $D_{a} = e^{\mu}_{\:\:a} D_{\mu}$ are ordinary
covariant derivatives, and indices in parentheses are to be
symmetrized completely, whilst those in square brackets are to be
anti-symmetrized, all with unit weight. Note in particular for $m
= 0$, $ f_{a}^{(1,n)\, \lh \mu_{1}...\mu_{n} \right.}\, e^{\left.
\mu_{n+1} \right)\, a}
  \, =\, 0.$
In a certain sense these equations represent a square root of
the generalized Killing equations, although they only provide
sufficient, not necessary conditions for obtaining solutions.
Having found $\Th$ we can then reconstruct the corresponding
$\cJ$. Eqs.(\ref{3.11.2}) partly solve $f^{(m+1,n-1)}$ in terms of
$f^{(m-1,n)}$ and only that part of $f^{(m+1,n-1)}$ is solved
which is symmetrized in one flat index and all $(n-1)$ curved
indices. On the other hand eqs.(\ref{3.11.2}) do not automatically
imply that $f^{(m+1,n-1)}$ is completely anti-symmetric in the
first $(m+1)$ indices. Imposing that condition on
eqs.(\ref{3.11.2}) one finds a new set of equations which are
precisely the generalized Killing equations for that part of
$f^{(m+1,n-1)}$ which was {\em not} given in terms of
$f^{(m-1,n)}$, and which should still be solved for. This is the
part of $f^{(m+1,n-1)}$ which is anti-symmetrized in one curved
index and all $(m+1)$ flat indices. Hence eqs.(\ref{3.11.2})
clearly have advantages over the generalized Killing equations
(\ref{3.6}). In order to find the constant of motion corresponding
to a Killing tensor of rank $n$,

\begin{equation}
\cD_{\lh \mu_{n+1} \rd}\, J^{(n)}_{\ld \mu_{1} ... \mu_{n}
\rh}\, =\, 0, \label{3.11.2.2}
\end{equation}
with ${ \cD_{\mu}}$ given by
\begin{equation}
{ \cD_{\mu}\ }  =  { \partial_{\mu}\, \, +\,
   \Gam_{\mu\nu}^{\:\:\:\:\:\lb}\, \Pi_{\lb}\, \dd{}{\Pi_{\nu}}\, +\,
   \og_{\mu\:\:b}^{\:\:a}\, \ps^{b}\, \dd{}{\ps^{a}}, } \\
\end{equation}
one has to solve the hierarchy of equations (\ref{3.6})
for $(J^{(n-1)}, ... , J^{(0)})$ and add the terms, as in
expression (\ref{3.5}). Having a solution
$f_{a_{1}...a_{m}}^{(m,n)\, \mu_{1}...\mu_{n} }$ of the equation $
f_{a_{1}...a_{m}}^{(m,n)\, \lh \mu_{1}...\mu_{n} \rd}\, e^{\ld
\mu_{n+1} \rh
   a_{1}}\, =\, 0,
\label{3.11.2.3} $ then we generate at least part of the
components $f_{a_{1}...a_{m+2\ag}}^{(m+2\ag, n-\ag)\,\mu_{1} ...
\mu_{n-\ag}}$ for $\ag = 1, ..., n$ by mere differentiation.

\section{New Supersymmetries and $f$-symbols}\label{sec-new-symm}

The constants of motion generate infinitesimal
transformations of the co-ordinates leaving the equations of
motion invariant: $ \del x^{\mu}\, =\, \del \ag\, \left\{ x^{\mu},
\cJ \right\}, \del \ps^{a}\, =\, \del \ag\, \left\{ \ps^{a}, \cJ
\right\}, \label{4.1} $
with $\del \ag$ the infinitesimal parameter of the transformation.
The theory might admit other (non-generic) supersymmetries \cite{Gibbons}
of the type $ \del x^{\mu}\, =\, -i \eps\, f^{\mu}_{\:\:a}\,
\ps^{a}\,$
with corresponding supercharges
\begin{equation}
{\cal Q}_{f}\, =\, -i \eps\, f^{\mu}_{\:\:a}\, \ps^{a}\, +\,
\frac{i}{3!}\, c_{abc}(x)\, \ps^{a} \ps^{b} \ps^{c}, \label{4.5}
\end{equation}
provided the tensors $f^{\mu}_{\:\:a}$ and $c_{abc}$ satisfy the differential constraints
\begin{equation}
D_{\mu}\, f_{\nu a}\, +\, D_{\nu}\, f_{\mu a}\, =\, 0, \label{4.6}
\end{equation}
\begin{equation}
D_{\mu}\, c_{abc}\, =\, - \lh R_{\mu\nu ab}\, f^{\nu}_{\:\:c}\,
+\,
   R_{\mu\nu bc}\, f^{\nu}_{\:\:a}\, +\, R_{\mu\nu ca}\, f^{\nu}_{\:\:b}\, \rh
{}. \label{4.7}
\end{equation}
One now obtains the following algebra of Poisson-Dirac brackets of
the conserved charges $\cQ_{i}$:
\begin{equation}
\left\{ {\cal Q}_{i}, {\cal Q}_{j} \right\}\, =\, -\, 2i\, Z_{ij},
\label{4.9}
\end{equation}
with
\begin{equation}
Z_{ij}\, =\, \frac{1}{2}\, K_{ij}^{\mu\nu}\, \Pi_{\mu} \Pi_{\nu}\,
+\,
             I^{\mu}_{ij}\, \Pi_{\mu}\, +\, G_{ij},
\label{4.10}
\end{equation}
where
\begin{equation}
\begin{array}{lll}
K^{\mu\nu}_{ij}  & = & \displaystyle{ \frac{1}{2}\, \lh f_{i\,
a}^{\mu} f_{j}^{\nu a}
+ f_{i\, a}^{\nu} f_{j}^{\mu a} \rh, } \\
  &  &  \\
\end{array}
\label{4.11}
\end{equation}
showing that $K_{ij\, \mu\nu}$ is a symmetric Killing
tensor of 2nd rank : $ D_{\lh \lb \rd}\, K_{ij\, (\ld \mu\nu \rh}
\, =\, 0,$ whilst $I^{\mu}_{ij}$ is the corresponding Killing vector and
$G_{ij}$ the corresponding Killing scalar.

In order to study the properties of the new supersymmetries, it is
convenient to introduce the 2nd rank tensor $ f_{\mu\nu}\, =\,
f_{\mu a}\, e^{\:\:a}_{\nu},$
which will be referred to as the {\em f-symbol} \cite{Gibbons}.
The defining relation (\ref{4.6}) implies
\begin{equation}
D_{\nu}\, f_{\lb \mu}\, +\, D_{\lb}\, f_{\nu \mu}\, =\, 0.
\label{5.1}
\end{equation}
It follows that the $f$-symbol is divergence-less on its
first index $ D_{\nu}\, f^{\nu}_{\:\:\mu}\,=\, 0.
$ and by contracting of eq.(\ref{5.1}) one finds $ D_{\nu}\,
f_{\mu}^{\:\:\nu}\, =\, - \partial_{\mu}\, f_{\nu}^{\:\:\nu}.$
Hence the divergence on the second index vanishes if and only if
the trace of the $f$-symbol is constant:
\begin{equation}
D_{\nu}\, f_{\mu}^{\:\:\nu}\, =\, 0 \hspace{1em} \Leftrightarrow
\hspace{1em}
          f_{\mu}^{\:\:\mu}\, =\, const.
\label{5.4}
\end{equation}
If the trace is constant, it maybe subtracted from the $f$-symbol
without spoiling condition (\ref{5.1}). It follows, that in this
case one may without loss of generality always take the constant
equal to zero and hence $f$ to be traceless. The symmetric part of
the $i$th $f$-symbol is the tensor
\begin{equation}
S_{\mu\nu}\, \equiv\, K_{i0\, \mu\nu}\, =\, \frac{1}{2}\, \lh
f_{\mu\nu} +
   f_{\nu\mu} \rh,
\label{5.4.1}
\end{equation}
which satisfies the generalized Killing equation $
D_{\left( \mu \right.}\, S_{\left. \nu \lb \right)}\, =\, 0.$
We can also construct the anti-symmetric part
\begin{equation}
B_{\mu\nu}\, =\, - B_{\nu\mu}\,
             =\, \frac{1}{2}\, \lh f_{\mu\nu} - f_{\nu\mu} \rh.
\label{5.7}
\end{equation}
which obeys the condition $ D_{\nu}\, B_{\lb\mu}\, +\,
D_{\lb}\, B_{\nu\mu}\, =\, D_{\mu}\, S_{\nu\lb}. \label{5.8} $ It
follows, that if the symmetric part does not vanish and is not
covariantly constant, then the anti-symmetric part $B_{\mu\nu}$ by
itself is {\em not} a solution of eq.(\ref{5.1}). But by the same
token the anti-symmetric part of $f$ can not vanish either, hence
$f$ can be completely symmetric only if it is covariantly
constant.\\
Anti-symmetric $f$-symbols, Killing-Yano tensors $f_{\mu
\nu}$ found by Penrose and Floyd \cite{PF}, and their corresponding
Killing-tensors have been studied extensively in
refs.\cite{Carter,CMcL,Visi1,Visi2,Visi3} in the related context of
finding solutions of the Dirac-equation in non-trivial curved
space-time.

\section{$f$-symbols in Robertson-Walker space-times}\label{sec-f-symb}

We consider space-time to be $\R\times\Sigma$, where
$\R$ represents the time direction and $\Sigma$ is a homogeneous and
isotropic three-manifold (maximally
symmetric space), with metric of the form
\begin{equation}
ds^2 = dt^2 - a^2(t)\gamma_{ij}(u)\d u^i\d u^j\ .\label{8.1}
\end{equation}
Here $t$ is the timelike coordinate, and $(u^1, u^2, u^3)$ are the
coordinates on $\Sigma$; $\gamma_{ij}$ is the maximally symmetric
metric on $\Sigma$. The function $a(t)$ is known as the
scale factor and the coordinates used here,
in which the metric is free of cross terms $\d t\,\d u^i$ and the
spacelike components are proportional to a single function of $t$, are
known as comoving coordinates.

Since the maximally symmetric metrics obey
$^{(3)}R_{ijkl} = k(\gamma_{ik}\gamma_{jl}
-\gamma_{il}\gamma_{jk})$,
where $k$ is some constant, the Ricci tensor is
$^{(3)}R_{jl} = 2k\gamma_{jl}$
and the metric on $\Sigma$ can be put in the form
\begin{equation}
  d\sigma^2 = \gamma_{ij}\d u^i\,\d u^j =
  e^{2\beta(r)}\d r^2 + r^2(\d\theta^2 +
  \sin^2\theta\,\d\phi^2)\label{8.4}
\end{equation}
we obtain the following metric on space-time:
\begin{equation}
  ds^2 = dt^2 - a^2(t)\left[{{\d r^2}\over{1-kr^2}}
  - r^2(\d\theta^2 +\sin^2\theta\,\d\phi^2)\right]\ .\label{8.7}
\end{equation}
with $k=-1,0,1$, known as Robertson-Walker metric.

We apply the results obtained previously to show that in a stationary
case $a(t)=const$ of this metric we obtain solutions
of (\ref{5.1}) which are not just Killing - Yano tensors of order two, also
investigated in \cite{hall, how, HowColl}.
In the stationary case and $k\neq0$ we have seven Killing vector fields:
\begin{eqnarray}\label{killing_vectors}
\vec{\bf \xi}^{(1)} &=&
-r\sqrt{1-kr^2}\sin(\theta)\frac\partial{\partial\theta}+
\frac1{1-kr^2}\cos(\theta)\frac\partial{\partial r}\nonumber\\
\vec{\bf \xi}^{(2)} &=&
\cos(\theta)\cos(\phi)\frac\partial{\partial\theta}-
\frac1{\sqrt{1-kr^2}}\sin(\theta)\cos(\phi)\frac\partial{\partial r}\nonumber\\
\vec{\bf \xi}^{(3)} &=&
r\sqrt{1-kr^2}\cos(\theta)\sin(\phi)\frac\partial{\partial\theta}+
\frac1{\sqrt{1-kr^2}}\sin(\theta)\sin(\phi)\frac\partial{\partial r}\nonumber\\
&+&r\sqrt{1-kr^2}\sin(\theta)\cos(\phi)\frac\partial{\partial\phi}\\
\vec{\bf \xi}^{(4)} &=& r^2\cos(\phi)\frac\partial{\partial\theta}-
r^2\sin(\theta)\cos(\theta)\sin(\phi)\frac\partial{\partial\phi}\nonumber\\
\vec{\bf \xi}^{(5)} &=& r^2\sin(\phi)\frac\partial{\partial\theta}+
r^2\sin(\theta)\cos(\theta)\cos(\phi)\frac\partial{\partial\phi}\nonumber\\
\vec{\bf \xi}^{(6)} &=&
r^2\sin^2(\theta)\frac\partial{\partial\phi}\nonumber\\
\vec{\bf \xi}^{(7)} &=& \frac\partial{\partial t}\nonumber
\end{eqnarray}
We obtain from (\ref{5.1}) the following three independent solutions
\begin{eqnarray}
f^{(1)}_{tt}&=&1\label{f-symb-sol-1}\\
f^{(2)}_{rt}&=&\frac{\cos(\theta)}{\sqrt{1-kr^2}},\ \ \ f^{(2)}_{\theta t}=-r\sqrt{1-kr^2}\sin(\theta)\label{f-symb-sol-2}\\
f^{(3)}_{\varphi t}&=&r^2\sin^2(\theta)\label{f-symb-sol-3}
\end{eqnarray}
and other four antisymmetric solutions (Killing-Yano tensors) \cite{how, HowColl},
\begin{eqnarray}\label{ky}
Y_{r\theta}^{(1)}&=&\frac{r\cos{\varphi}}{\sqrt{1-kr^2}},\
Y_{r\varphi}^{(1)}=-\frac{r\sin{\theta}\cos{\theta}\sin{\varphi}}{\sqrt{1-kr^2}},\
Y_{\theta\varphi}^{(1)}=r^2\sqrt{1-kr^2}\sin^2{\theta}\sin{\varphi}\nonumber\\
Y_{r\theta}^{(2)}&=&\frac{r\sin{\varphi}}{\sqrt{1-kr^2}},\ Y_{r\varphi}^{(2)}=\frac{r\sin{\theta}\cos{\theta}\cos{\varphi}}{\sqrt{1-kr^2}},\
Y_{\theta\varphi}^{(2)}=-r^2\sqrt{1-kr^2}\sin^2{\theta}\cos{\varphi}\\
Y_{r\varphi}^{(3)}&=&\frac{r\sin^2{\theta}}{\sqrt{1-kr^2}},\ Y_{\theta\varphi}^{(3)}=r^2\sqrt{1-kr^2}\sin{\theta}\cos{\theta}\nonumber\\
Y_{\theta\varphi}^{(4)}&=&r^3\sin{\theta}.\nonumber
\end{eqnarray}
By taking the symmetric parts (\ref{5.4.1}) for solutions
$f^{(1)}$, $f^{(2)}$ and $f^{(3)}$, we obtain the following three Killing
tensors:
\begin{eqnarray}
K^{(1)}_{tt}&=&1\\
K^{(2)}_{tr}&=&\frac 12\frac{\cos(\theta)}{\sqrt{1-kr^2}},\ \ \ K^{(1)}_{t\theta}=-\frac 12r\sqrt{1-kr^2}\sin(\theta)\\
K^{(3)}_{t\varphi}&=&\frac 12r^2\sin^2(\theta)
\end{eqnarray}
From (\ref{4.11}), (\ref{f-symb-sol-1}), (\ref{f-symb-sol-2}),
(\ref{f-symb-sol-3}) we obtain six more Killing tensors
\begin{eqnarray}
K^{(4)}_{tt}&=&1\\
K^{(5)}_{tr}&=&\frac12\frac{\cos(\theta)}{\sqrt{1-kr^2}},\ \
K^{(5)}_{t\theta}=-\frac{1}2r\sqrt{1-kr^2}\sin(\theta)\\
K^{(6)}_{t\varphi}&=&\frac12r^2\sin^2(\theta)\\
K^{(7)}_{rr}&=&\frac{\cos^2(\theta)}{(1-kr^2)},\ \
K^{(7)}_{\theta\theta}=r^2(1-kr^2)\sin^2(\theta),\ \
K^{(7)}_{r\theta}=-r \sin(\theta)\cos(\theta)\\
K^{(8)}_{\varphi\varphi}&=&r^4\sin^4(\theta)\ \ \\
K^{(9)}_{r\varphi}&=&\frac12\frac{r^2}{\sqrt{1-kr^2}}\sin^2(\theta)\cos(\theta),\ \
K^{(9)}_{\theta\varphi}=-\frac12r^3\sqrt{1-kr^2}\sin^3(\theta)
\end{eqnarray}
We observe that the Killing tensors obtained from
$f^{(1)}$, $f^{(2)}$ and $f^{(3)}$ are reducible
\begin{eqnarray}
K^{(1)}_{\mu\nu} &=& \xi^{(7)}_\mu\xi^{(7)}_\nu\\
K^{(2)}_{\mu\nu} &=&
\frac12\left(\xi^{(1)}_\mu\xi^{(7)}_\nu + \xi^{(1)}_\nu\xi^{(7)}_\mu\right)\\
K^{(3)}_{\mu\nu} &=&
\frac12\left(\xi^{(6)}_\mu\xi^{(7)}_\nu + \xi^{(6)}_\nu\xi^{(7)}_\mu\right)\\
K^{(4)}_{\mu\nu} &=& \xi^{(7)}_\mu\xi^{(7)}_\nu\\
K^{(5)}_{\mu\nu} &=&
\frac12\left(\xi^{(1)}_\mu\xi^{(7)}_\nu + \xi^{(1)}_\nu\xi^{(7)}_\mu\right)\\
K^{(6)}_{\mu\nu} &=&
\frac12\left(\xi^{(6)}_\mu\xi^{(7)}_\nu + \xi^{(6)}_\nu\xi^{(7)}_\mu\right)\\
K^{(7)}_{\mu\nu}&=&\xi^{(1)}_\mu\xi^{(1)}_\nu\\
K^{(8)}_{\mu\nu}&=&\xi^{(6)}_\mu\xi^{(6)}_\nu\\
K^{(9)}_{\mu\nu}&=&\frac12\left(\xi^{(1)}_\mu\xi^{(6)}_\nu+\xi^{(1)}_\nu\xi^{(6)}_\mu\right)
\end{eqnarray}
as well as Killing tensors \cite{how, HowColl} constructed from Killing-Yano tensors (\ref{ky}),
where reducible means that it can be written as a linear
combination of the metric and symmetrized products of Killing vectors, i.e.
$$
K_{\mu\nu}=a_0g_{\mu\nu}+\sum\limits_{I=1}^{N}\sum\limits_{J=1}^{N}a_{IJ}\xi_{(\mu}^{(I)}\xi_{\nu)}^{(J)}
$$
where $\xi^{(I)}$ for $I = 1\dots N$ are the Killing vectors admitted by the manifold and $a_0$ and
$a_{IJ}$ for $1\leq I\leq J\leq N$ are constants,
the quadratic constant of motion associated
with a reducible Killing tensor simply being a linear combination of
existing first integrals.

The case investigated does not
correspond to any of the special cases \cite{HauMal} as admitting non-reducible
Killing tensors of order two that exist in spherically symmetric static space-times.

\section{Conclusions}\label{sec-concl}

In this paper a spinning particle model was investigated and was
shown, that in a stationary case of Robertson-Walker space-time, there
exists  a class of new structures called {\em f-symbols} which can
generate reducible Killing tensors and supersymmetry algebras.
As to our knowledge a curved space-time possessing such structures
have not been exemplified until now.
Further implications of {\em f-symbols} in the context of
Dirac-type operators \cite{CMcL,Visi1,Visi2,Visi3,Klishevich} and
geometric duality \cite{Ri1,D1,D2,D3} are under investigations
\cite{PT}.

\subsection*{Acknowledgments}

The authors are grateful to Mihai Visinescu for valuable
suggestions and discussions. This work is supported by M.E.C,
NUCLEU - LAPLACE 03-170601.7.

\end{document}